\documentclass[preprint,showpacs,preprintnumbers,amsmath,amssymb,nofootinbib]{revtex4}
\usepackage{booktabs}
\usepackage{mathrsfs}
\usepackage{epsfig}
\usepackage{graphicx}
\usepackage{dcolumn}
\usepackage{bm}
\usepackage{amsmath}
\usepackage{slashed}
\usepackage{multirow}
\usepackage{setspace}

\let\jnfont=\rm
\def\NPB#1,{{\jnfont Nucl.\ Phys.\ B }{\bf #1},}
\def\PLB#1,{{\jnfont Phys.\ Lett.\ B }{\bf #1},}
\def\EPJC#1,{{\jnfont Eur.\ Phys.\ Jour.\ C }{\bf #1},}
\def\PRD#1,{{\jnfont Phys.\ Rev.\ D }{\bf #1},}
\def\PRL#1,{{\jnfont Phys.\ Rev.\ Lett.\ }{\bf #1},}
\def\MPLA#1,{{\jnfont Mod.\ Phys.\ Lett.\ A }{\bf #1},}
\def\JPG#1,{{\jnfont J.\ Phys.\ G}{\bf #1},}
\def\CTP#1,{{\jnfont Commun.\ Theor.\ Phys.\ }{\bf #1},}
\def\ZPC#1,{{\jnfont Z.\ Phys.\ C }{\bf #1},}
\def\JHEP#1,{{\jnfont JHEP \ }{\bf #1},}
\def\Rv{\not{\hbox{\kern-1pt $R$}}}
\def\p{\not{\hbox{\kern-3pt $p$}}}

\begin{document}

\title{\boldmath SUSY effects in $R_b$: revisited under current experimental constraints}
\author{ Wei Su$^{1}$}
\author{ Jin Min Yang$^{1,2}$}

\affiliation{$^1$ Institute of Theoretical Physics, Academia Sinica, Beijing 100190, China\\
$^2$ Department of Physics, Tohoku University, Sendai 980-8578, Japan
   \vspace*{1.5cm} }%

\date{\today}

\begin{abstract}

In this note we revisit the SUSY effects in $R_b$ under
current experimental constraints including the LHC Higgs data, the $B$-physics
measurements, the dark matter relic density and direct detection limits, as well as
the precision electroweak data. We first perform a scan to figure out the currently
allowed parameter space and then display the SUSY effects in
$R_b$. We find that although the SUSY parameter space has been severely
restrained by current experimental data, both the general MSSM and the natural-SUSY
scenario can still alter $R_b$ with a
magnitude sizable enough to be observed at future $Z$-factories (ILC, CEPC, FCC-ee,
Super $Z$-factory) which produce $10^9-10^{12}$ $Z$-bosons.
To be specific, assuming a precise measurement $\delta R_b = 2.0 \times 10^{-5}$  at FCC-ee,
we can probe a right-handed stop up to 530 GeV through chargino-stop loops,
probe a sbottom to 850 GeV through neutralino-sbottom loops
and a charged Higgs to 770 GeV through the Higgs-top quark loops
for a large $\text{tan}\beta$.
The full one-loop SUSY correction to $R_b$ can reach $1 \times 10^{-4}$
in natural SUSY and $2 \times 10^{-4}$ in the general MSSM.

\end{abstract}

\maketitle
\begin{spacing}{1.2}

\section{Introduction}
\label{sec:introduction}
After the discovery of the 125 GeV Higgs boson \cite{Chatrchyan:2012ufa,Aad:2012tfa},
the primary task of the LHC is to hunt for new physics beyond the Standard Model (SM).
Among various extensions of the SM, the low energy supersymmetry (SUSY) is the most
appealing candidate\footnote{Confronted with the 125 GeV Higgs mass, the minimal SUSY
model (MSSM) has a little fine-tuning while the next-to-minimal SUSY model is more
favored \cite{mssm-nmssm}.} since it can solve the gauge hierarchy problem,
naturally explain
the cosmic cold dark matter and achieve the gauge coupling unification. The search for
SUSY has long been performed both directly and indirectly. On the one hand, the colliders
have directly searched for the sparticle productions. On the other hand, SUSY
effects have been probed indirectly through precision measurements of some low energy
observables.

$R_b \equiv \Gamma(Z\to\bar bb)/\Gamma(Z \to hadrons)$
is a famous observable which is sensitive to new physics beyond the SM \cite{Rb}.
So far the most precise experimental value $R_b^{\rm exp}= 0.21629\pm0.00066$ comes from the
LEP and SLC measurements \cite{ALEPH:2005ab}, while the SM prediction is
$R_b^{\rm SM}=0.21579$ \cite{Freitas:2014hra}.
The future $Z$-factories are expected to produce much more $Z$-bosons
than the LEP experiment. For example, $10^9$, $10^{10}$ and  $10^{12}$  $Z$-bosons
are expected to be produced respectively at the International Linear Colldier (ILC) \cite{ILC},
the Circular Electron-Positron Collider (CEPC) \cite{CEPC}, the Future Circular Collider (FCC-ee)
\cite{FCC-ee} and the Super $Z$-factory \cite{super-Z}.
This will allow for a more precise measurement of $R_b$ \cite{Fan:2014axa}
and help pin down the involved new physics effects.

The SUSY effects in $R_b$ were calculated and discussed many years ago \cite{Boulware:1991vp,Cao:2008rc,Garcia:1994wv,history}.
In this work we revisit these effects for two reasons:
(i) The current experiments, especially the LHC experiments, have severely restrained the
SUSY parameter space. It is intriguing to figure out the possible magnitude of
the SUSY effects in the currently allowed parameter space;
(ii) Given the possibility of some future  $Z$-factories like ILC, CEPC or FCC-ee, a more precise
measurement of $R_b$ will help reveal the SUSY effects although these effects may have already
been restrained to be rather small by current experiments. In order to know if the SUSY
effects are accessible in a future measurement of $R_b$, we must figure out their currently allowed
value.

This work is organized as follows.
In Sec.\ref{sec:formula}, we give a description of SUSY effects in $R_b$.
In Sec.\ref{sec:result}, we scan over the SUSY parameter space and display the
SUSY effects in the allowed parameter space.
Finally we give our conclusion in Sec.~\ref{sec:conclusion}.

\section{SUSY corrections to $R_b$}
\label{sec:formula}
\begin{figure}
  \centering
  \includegraphics[width=0.7\textwidth]{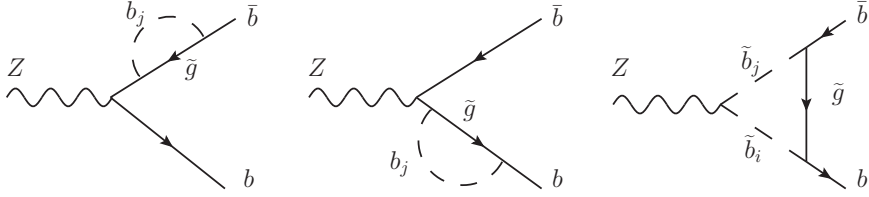}
  \vspace{-0.3cm}
  \caption{One-loop Feynman diagrams of gluino correction to $Z \to \bar b b$}
  \label{fig:gluino}
\end{figure}

\begin{figure}
  \begin{minipage}[t]{0.5\linewidth}
    \centering
    \includegraphics[width=0.9\textwidth]{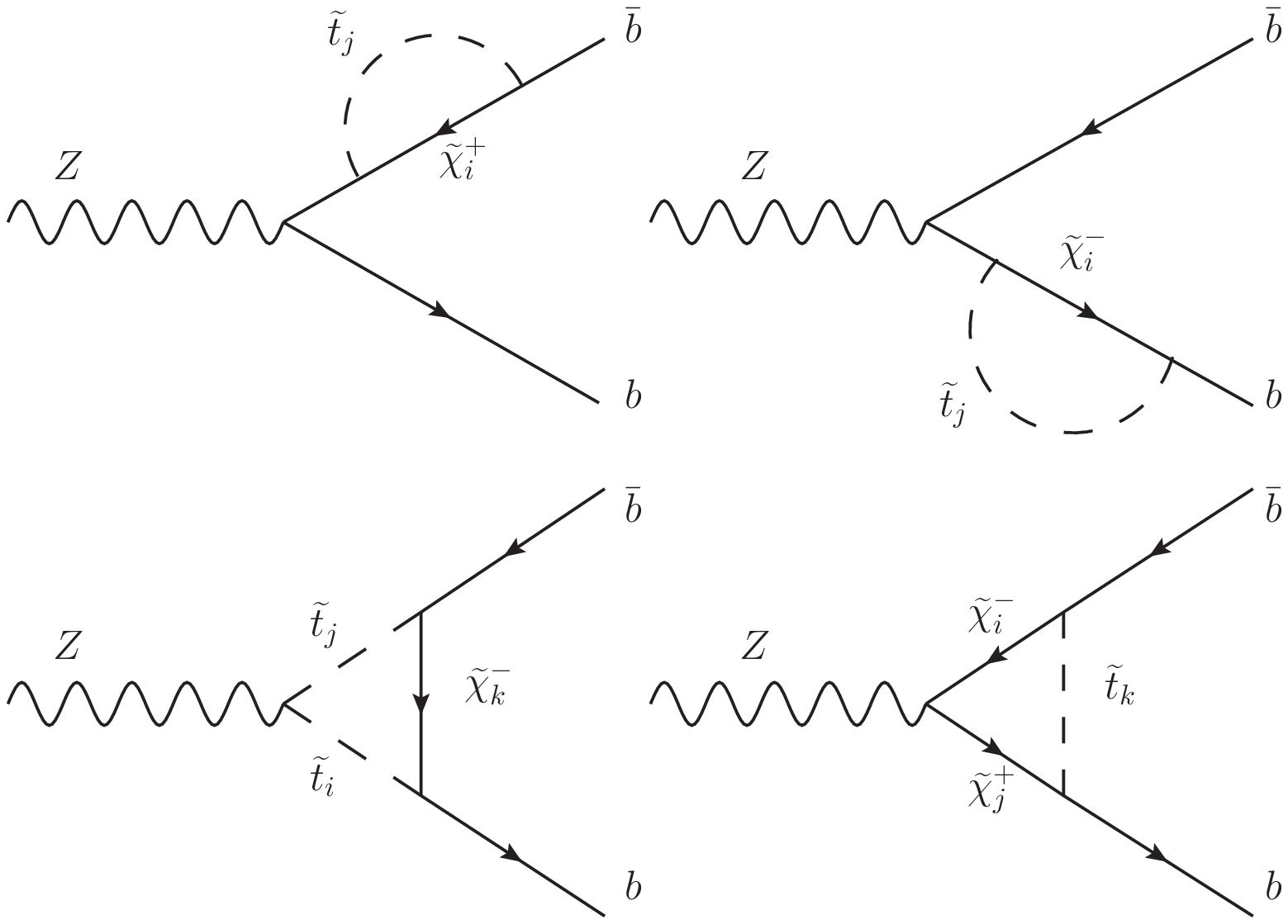}
   \vspace{-0.3cm}
   \caption{One-loop Feynman diagrams of \protect\\ chargino correction to $Z \to \bar b b$}
    \label{fig:chargino}
  \end{minipage}%
  \begin{minipage}[t]{0.5\linewidth}
    \centering
    \includegraphics[width=0.9\textwidth]{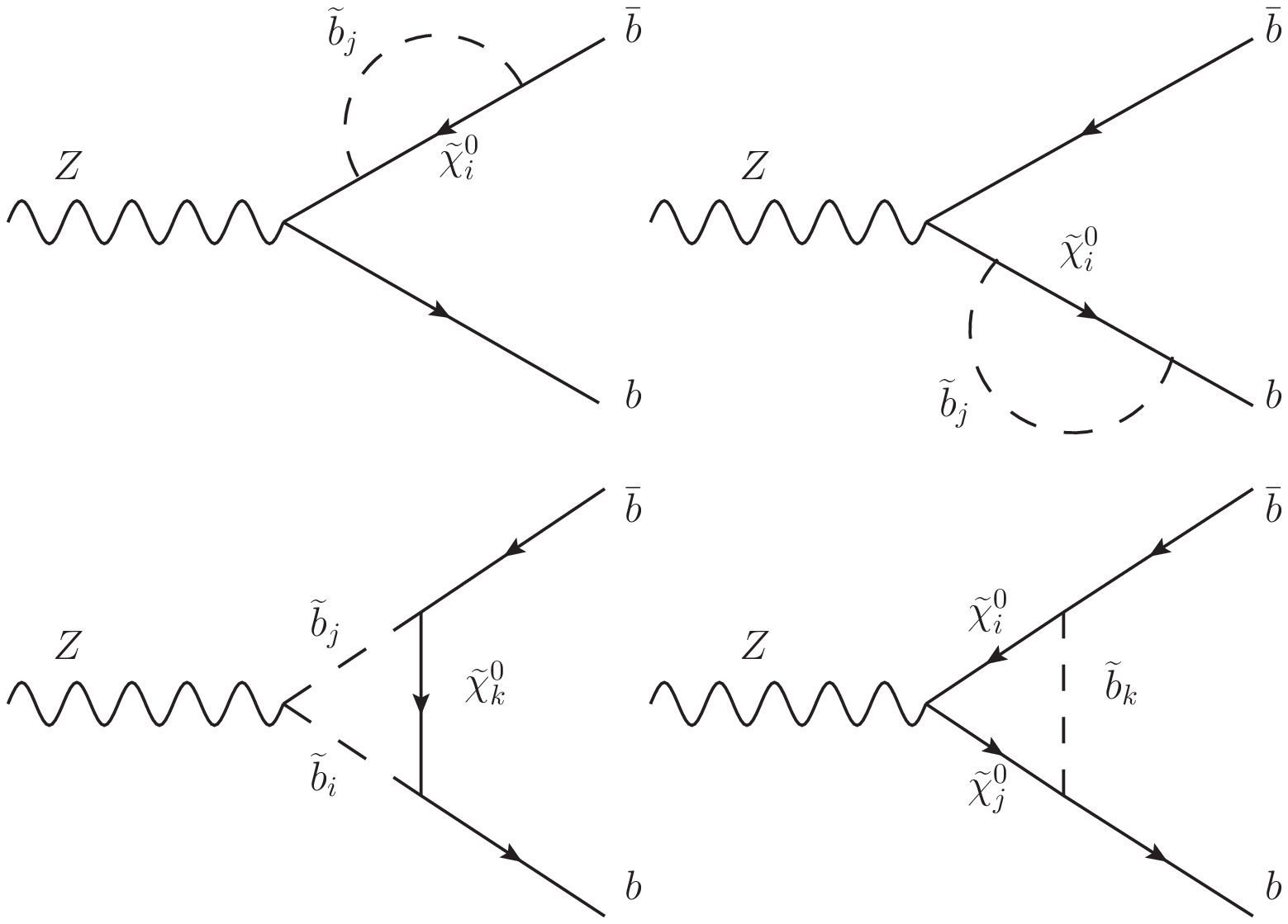}
    \vspace{-0.3cm}
    \caption{One-loop Feynman diagrams of \protect\\neutralino correction to $Z \to \bar b b$}
    \label{fig:neutralino}
  \end{minipage}
\end{figure}

\begin{figure}
  \begin{minipage}[t]{0.45\linewidth}
    \centering
    \includegraphics[width=0.9\textwidth]{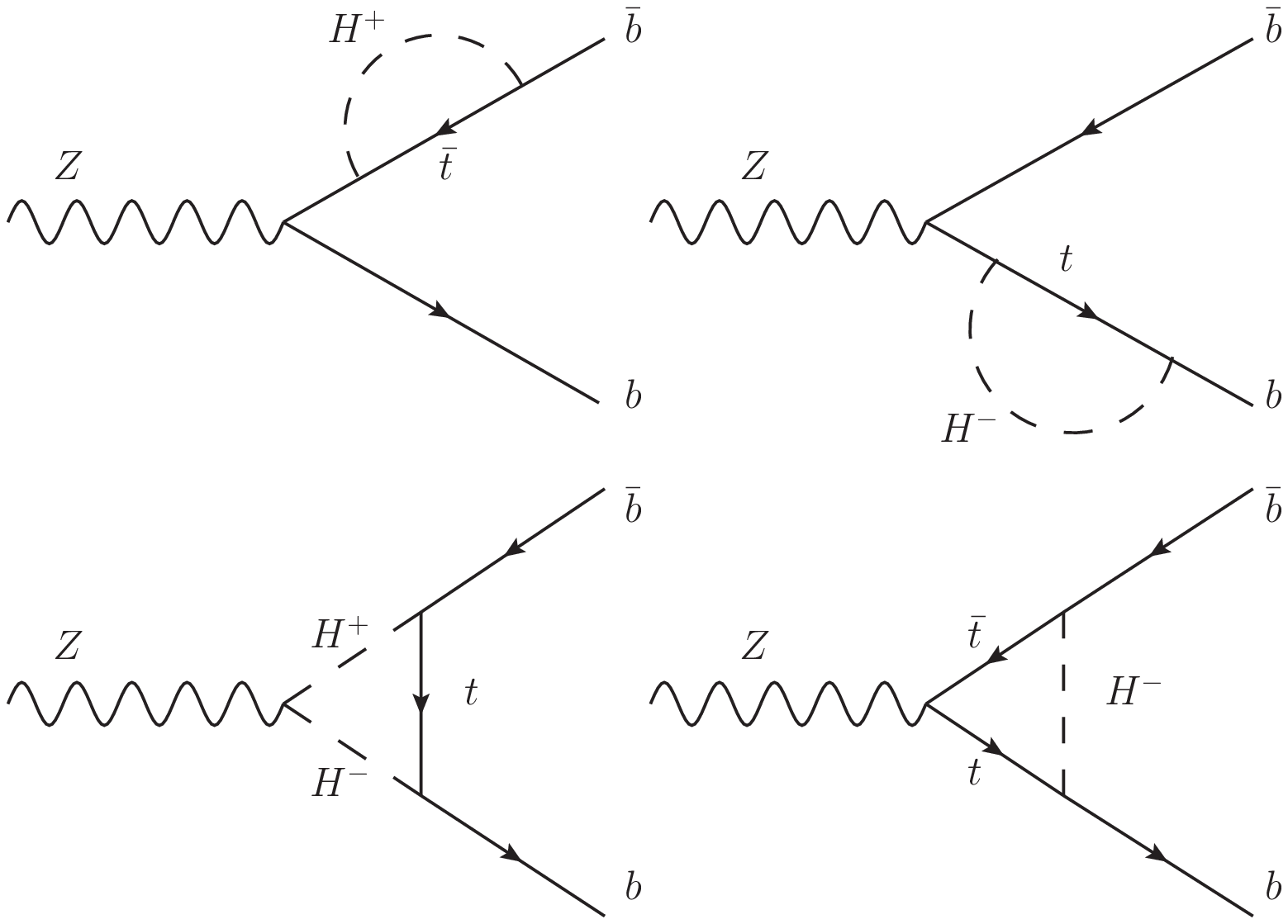}
    \vspace{-0.3cm}
    \caption{One-loop Feynman diagrams of \protect\\ charged Higgs correction to $Z \to \bar b b$}
    \label{fig:chargedhiggs}
  \end{minipage}
  \begin{minipage}[t]{0.45\linewidth}
    \centering
    \includegraphics[width=0.9\textwidth]{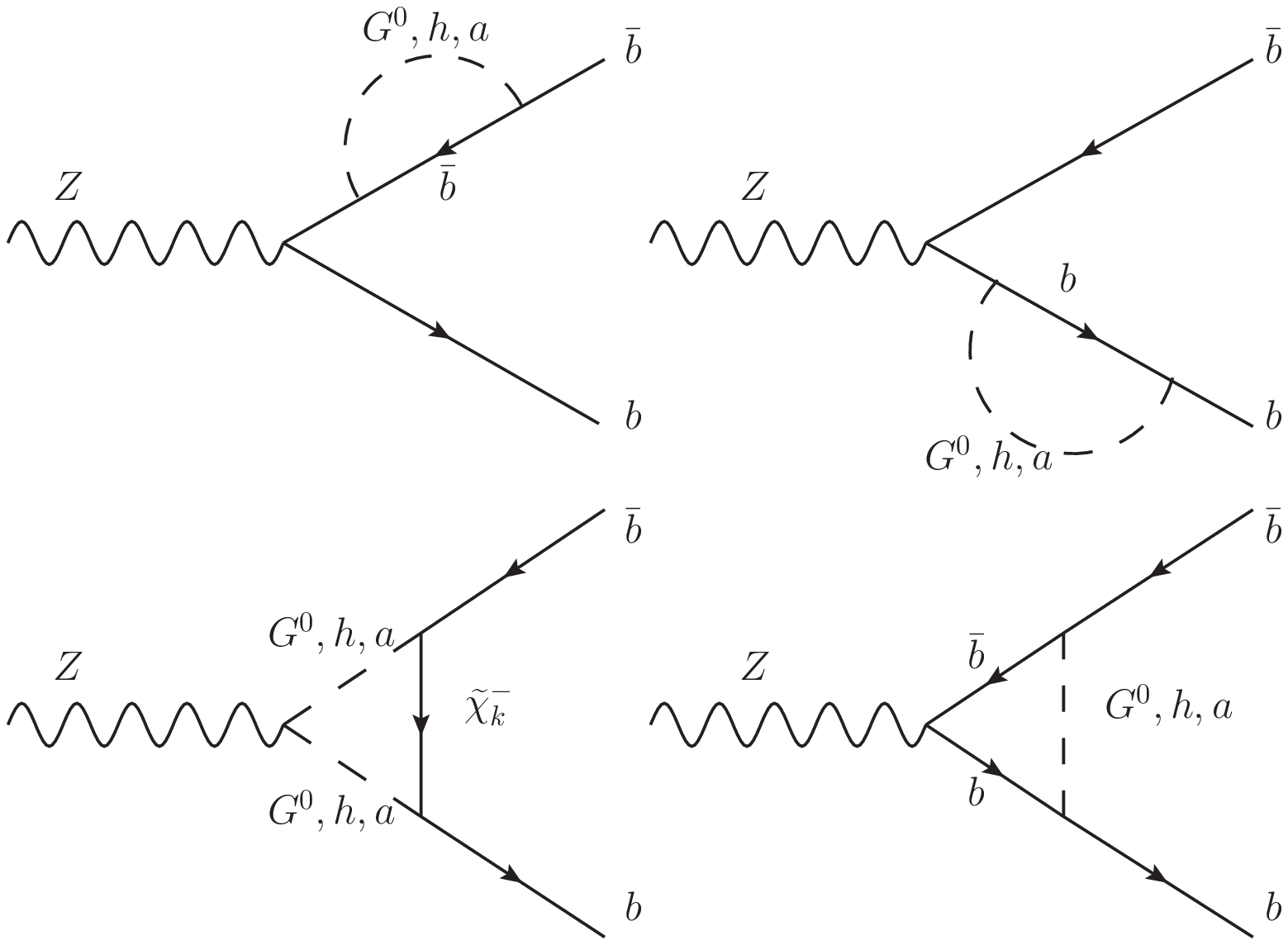}
    \vspace{-0.3cm}
    \caption{One-loop Feynman diagrams of \protect\\ neutral Higgs correction to $Z \to \bar b b$}
    \label{fig:neutralhiggs}
  \end{minipage}
\end{figure}

Since the SUSY effects in $R_b$ have been calculated in the literature \cite{Boulware:1991vp,Cao:2008rc},
here we only give a brief description.
The dominant SUSY effects in $R_b$ are from the vertex corrections to $Z\to b\bar{b}$, as shown in
Figs.~\ref{fig:gluino}-\ref{fig:neutralhiggs}.
These corrections come from the gluino loops, chargino loops, neutralino loops,
charged Higgs loops and neutral Higgs loops.

%
The one-loop SUSY correction to $R_b$ can be expressed as
\begin{equation}
 \label{eq:factRb}
 \delta R^{SUSY}_b \simeq \frac{R_b^{SM}(1-R_b^{SM})}{v^2_b(3-\beta^2) + 2a_b^2\beta^2}[v_b(3-\beta^2) \delta v_b+2a_b\beta^2\delta {a_b}],
 \end{equation}
where $v_b = 1/2-2sin\theta_w^2/3$ and $a_b = 1/2$ are respectively the vector and axial
vector couplings of tree-level $Zb\bar{b}$ interaction, $\beta = \sqrt{1-4m_{b}^2/m_Z^2}$ is the
velocity of bottom quark in $Z \to b\bar{b}$, and $\delta v_b$ and $\delta a_b$ are the
corresponding corrections defined as~\cite{Cao:2008rc,Bohm:1986rj,Hollik:1988ii}
 \begin{equation}\label{eq:delta abvb}
   \delta v_b = \frac{\delta g^b_L + \delta g^b_R}{2},\quad \delta a_b = \frac{\delta g^b_L - \delta g^b_R}{2}.
 \end{equation}
Here $\delta g^b_{\lambda}$ ($\lambda = L, R$) is give by
 \begin{equation}\label{eq:delta gLR}
 \delta g^b_{\lambda} = \Gamma_{f{\lambda}}(m_Z^2) - g_{\lambda}^{Zb\bar{b}}\Sigma_{b{\lambda}}(m_b^2),
 \end{equation}
where $\Gamma_{f{\lambda}}(m_Z^2)$ denotes the vertex loop contributions
and $\Sigma_{b{\lambda}}(m_b^2)$ is the counter term from the bottom quark self-energy.
We perform straightforward loop calculations and confirm the expressions in \cite{Cao:2008rc}.
The results can be expressed as
\begin{eqnarray}
 \Sigma_{b_\lambda}(p_b^2) &=&
    \frac{C_g}{(4\pi)^2} \biggl| g^{\bar{\psi}_j b \phi_i^\ast}_\lambda \biggr|^2
    ( B_0 + B_1 ) (p_b, m_{\phi_i}, m_{\psi_j}), \\
  \Gamma_{b_\lambda}(q^2) &=&
    -\frac{C_g}{(4\pi)^2} \Biggl\{
    \biggl( g_\lambda^{\bar{\psi}_j b \phi_k^\ast} \biggr)^*
    g_\lambda^{\bar{\psi}_i b \phi_k^\ast}
    \biggl[
    g_\lambda^{\bar{\psi}_j \psi_i Z} m_{\psi_i} m_{\psi_j} C_0
\nonumber \\
&&
    + g_{-\lambda}^{\bar{\psi}_j \psi_i Z}
    \biggl(-q^2 (C_{12} + C_{23}) - 2 C_{24} +  \frac{1}{2} \biggr)
    \biggr] (p_{\bar{b}}, p_{b}, m_{\psi_i}, m_{\phi_k}, m_{\psi_j})
\nonumber \\
&&
    -  \biggl( g^{\bar{\psi}_k b \phi_i^\ast}_\lambda \biggr)^*
    g^{\bar{\psi}_k b \phi_j^\ast}_\lambda
    g^{\phi_i^\ast \phi_j Z}
    2 C_{24}(p_{\bar{b}}, p_b, m_{\phi_j}, m_{\psi_k}, m_{\phi_i})
    \Biggr\},
\end{eqnarray}
where $C_g = 4/3$ for the gluino loops and $C_g=1$ for other loops,
and $B_0$, $B_1$ and $C_{12}$, $C_{23}$, $C_{24}$ are Passarino-Veltman functions ~\cite{Passarino:1978jh}.
The notation $(\phi,\psi)$ represents ($\tilde b, \tilde g$) for gluino loops,
($\tilde t, \tilde\chi^-$) for chargino loops, ($\tilde b, \tilde\chi^0$) for neutralino loops,
($H^-, t$) for charged Higgs loops and ($h/a/G^0$, $b$) for neutral Higgs loops.

In addition to $R_b$, we also show the SUSY effects in the forward-backward asymmetry
$A^b_{FB}$ in the decay $Z \to \bar{b}b$:
\begin{equation}
 \label{eq:abfb}
 \delta A^{b}_{FB}\big|_{SUSY} \simeq A^{b}_{FB}\big|_{SM}\big(\frac{v_b \delta v_b + a_b \delta a_b}{a_b v_b} - 2\frac{v_b(3-\beta^2) \delta v_b+2a_b\beta^2\delta {a_b}}{v^2_b(3-\beta^2) + 2a_b^2\beta^2}\big).
 \end{equation}
Its experimental value is $0.0992 \pm 0.0016$ from the LEP experiment \cite{ALEPH:2005ab}
while its SM prediction is $0.1032 \pm 0.0004$~\cite{Baak:2014ora}. In the future $Z$-factories,
this forward-backward asymmetry will be measured together with $R_b$, both of which will jointly
allow for a revelation of SUSY effects.

\section{Numerical calculations and results}
\label{sec:result}
\subsection{SUSY parameter space}
To clarify our numerical calculations we consider the general MSSM and the natural-SUSY scenario
\cite{natural-susy}. From the natural-SUSY results (the natural-SUSY parameter space is much
smaller than the general MSSM), we can acquire the more detailed characters of each kind of loops,
while from the general MSSM results we can obtain the more general size of SUSY loop effects.

For the natural-SUSY scenario, since in this scenario only the higgsino masses and the third-generation
squark masses are assumed to be light, while other sparticles are assumed to be
rather heavy and thus their effects in low energy observables are decoupled,
in our scan we fix the soft-breaking mass
parameters in the first two generation squark sector and the slepton sector
at 5 TeV, and assume $A_t = A_b$.
For the electroweak gaugino masses, inspired by the grand unification relation,
we take $M_1 : M_2 = 1 : 2$ and fix $M_2$ at 2 TeV.
The gluino mass is fixed at 2 TeV since it is supposed to be not too far
above TeV scale in natural-SUSY. Other parameters vary as follows
\begin{eqnarray}
&& 1 < \tan\beta < 60, 100~{\rm GeV}< \mu < 200 ~{\rm GeV}, ~ |A_t| < 3 ~{\rm TeV},   \nonumber\\
&& 100 ~{\rm GeV} < m_{Q_3}, m_{U_3} , m_{D_3}< 2 ~{\rm TeV}.
\end{eqnarray}
For the general MSSM,  assuming $A_t = A_b$ and $M_1 : M_2:M_3 = 1 : 2:6$, we scan over the
following parameter space
\begin{eqnarray}
&& 1 < \tan\beta < 60, 100~{\rm GeV}< \mu < 1000 ~{\rm GeV}, ~ |A_t| < 3 ~{\rm TeV},   \nonumber\\
&& 100 ~{\rm GeV} < m_{Q_3}, m_{U_3} , m_{D_3}< 2 ~{\rm TeV},100~{\rm GeV}< M_2 < 20000 ~{\rm GeV}.
\end{eqnarray}
In our scan we consider the following experimental constraints:
\begin{itemize}
  \item[(1)] The constraints on the Higgs sector from the LEP, Tevatron and LHC experiments.
We use the package HiggsBounds-4.0.0~\cite{Bechtle:2011sb} to implement these constraints.
  \item[(2)] The experimental constraints in $B$-physics. We require SUSY to satisfy
various $B$-physics bounds at $2\sigma$ level with SUSY FLAVOR v2.0~\cite{Crivellin:2012jv}, which includes
$B\to X_s\gamma$, $B_s\to \mu^+\mu^-$, $B^+\to\tau^+\nu$ and so on~\cite{Agashe:2014kda}.
 \item[(3)] The measurements of the precision electroweak observables.
The SUSY predictions of $\rho_l$, $sin^2\theta_{eff}^l$ and $m_W$ are required to be within the
2$\sigma$ ranges of the experimental values~\cite{ALEPH:2005ab}.
  \item[(4)] The dark matter constraints. We require the thermal relic density of the neutralino
dark matter to be below the 2$\sigma$ upper limit of the Planck value~\cite{Ade:2013zuv}
and require the dark matter-nucleon
spin-independent scattering scross section $\sigma_r^{SI}$
to satisfy the 95\% C.L. limits of LUX~\cite{Akerib:2013tjd}.
We also consider the limits of spin-dependent dark matter-nucleon cross section $\sigma_r^{SD}$
from the XENON100 experiment~\cite{Aprile:2013doa}.
The relic density, $\sigma_r^{SI}$ and $\sigma_r^{SD}$ are calculated with the code MicrOmega v2.4~\cite{Belanger:2010gh}.
\end{itemize}
About the mass bounds from the LHC direct searches, in natural SUSY the higgsinos have very weak
bounds because their pair productions only give missing energy and are rather difficult to detect
(a mono-jet or mono-$Z$ is needed in detection) \cite{mono-jet}, while for the stops the
right-handed one is weakly bounded (its mass can be as light as 210 GeV for higgsinos heavier
than 190 GeV) \cite{stop-bound}. When we display the numerical results, we will not show a sharp
LHC bound on stop or higgsino mass (we only consider the LEP bounds on stop and higgsinos).
For each surviving sample we calculate the correction to $R_b$ and display the
numerical results in the proceeding section.

\subsection{Numerical results of $R_b$ and $A^{b}_{FB}$}
\begin{figure}
  \centering
  \includegraphics[width=0.45\textwidth]{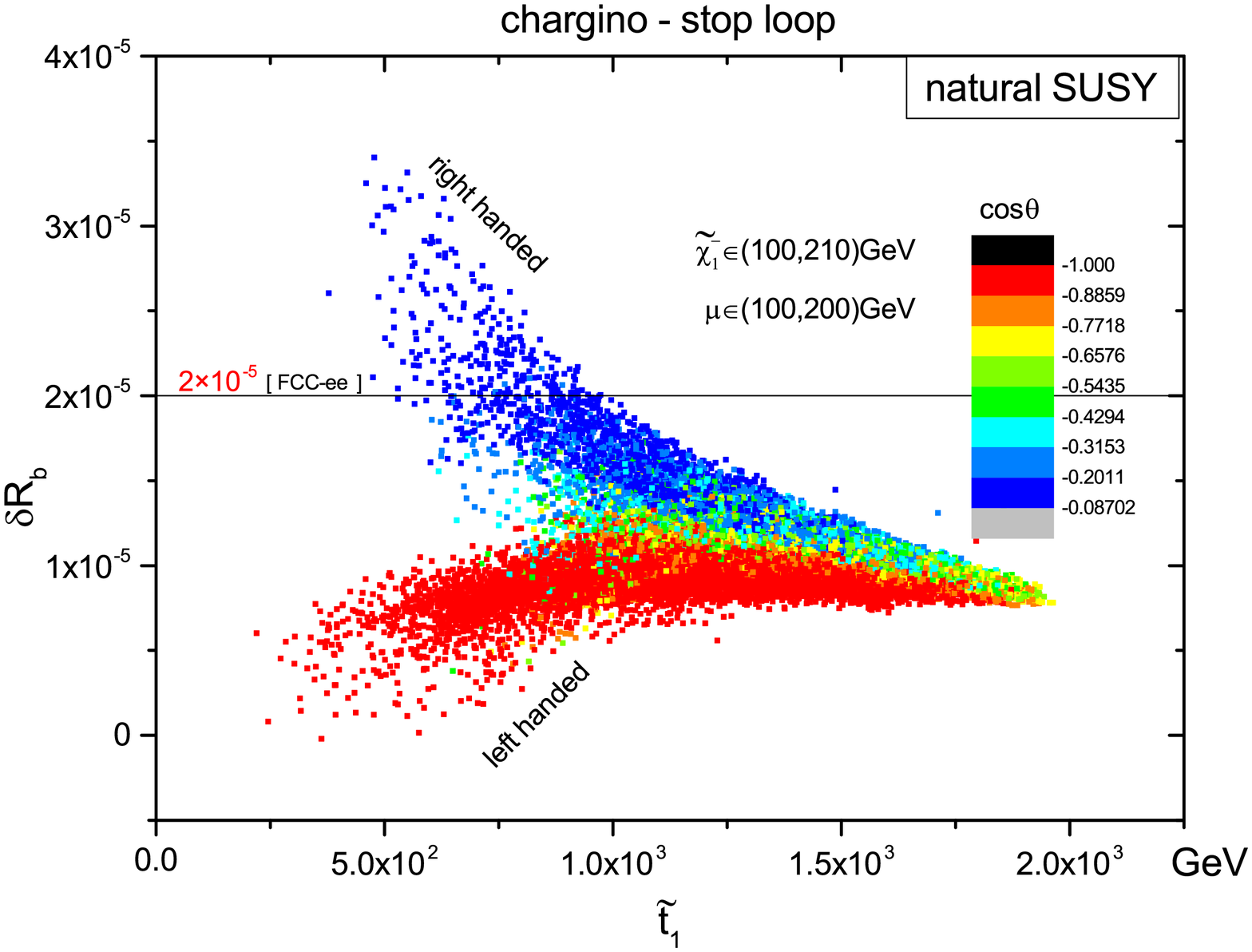}
  \includegraphics[width=0.45\textwidth]{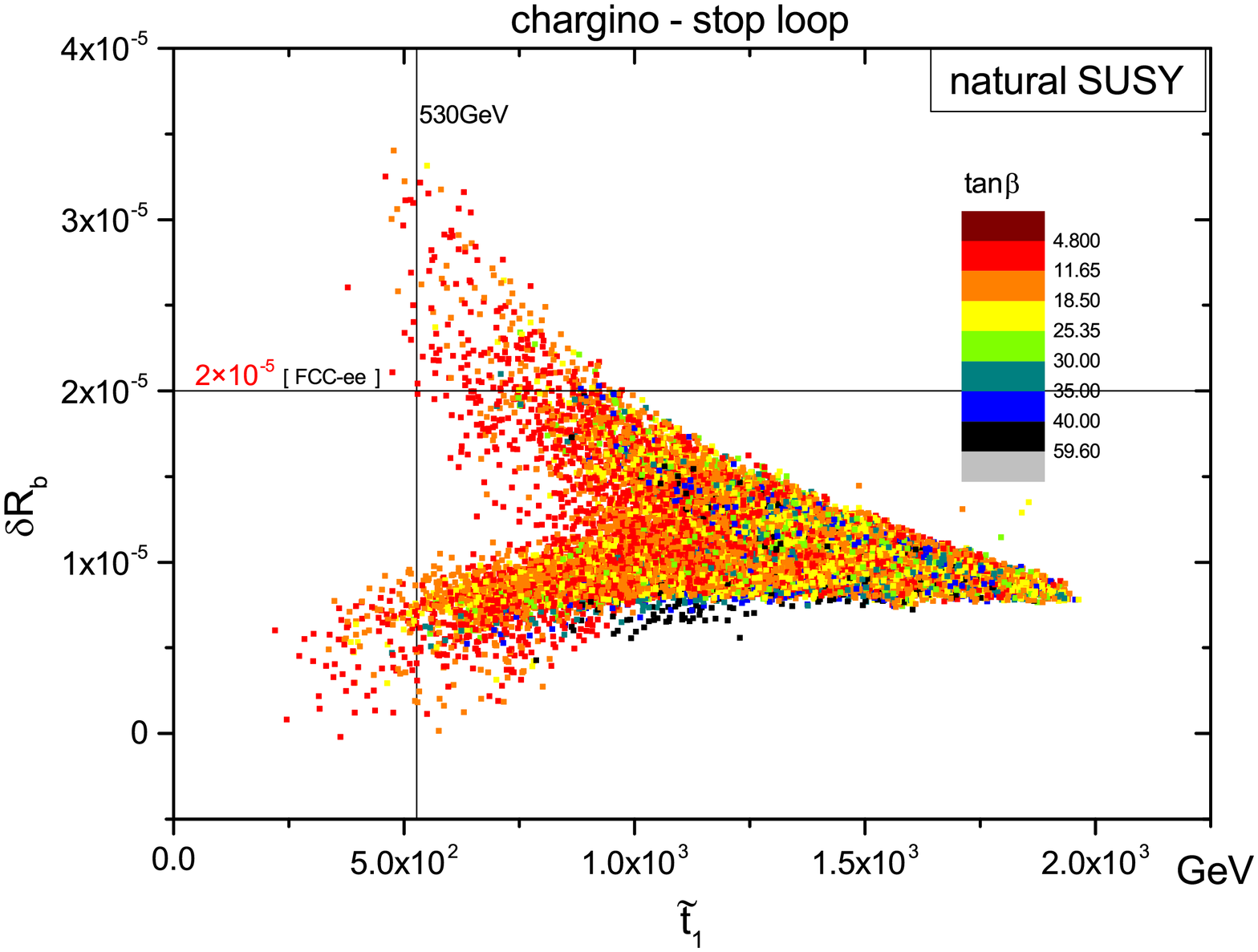}
   \vspace{-0.5cm}
  \caption{The scatter plots of the surviving samples of natural SUSY, showing the chargino one-loop effects in $R_b$.}
  \label{dia:chargino}
\end{figure}

\begin{figure}
  \begin{minipage}[t]{0.5\linewidth}
    \centering
    \includegraphics[width=1.0\textwidth,clip]{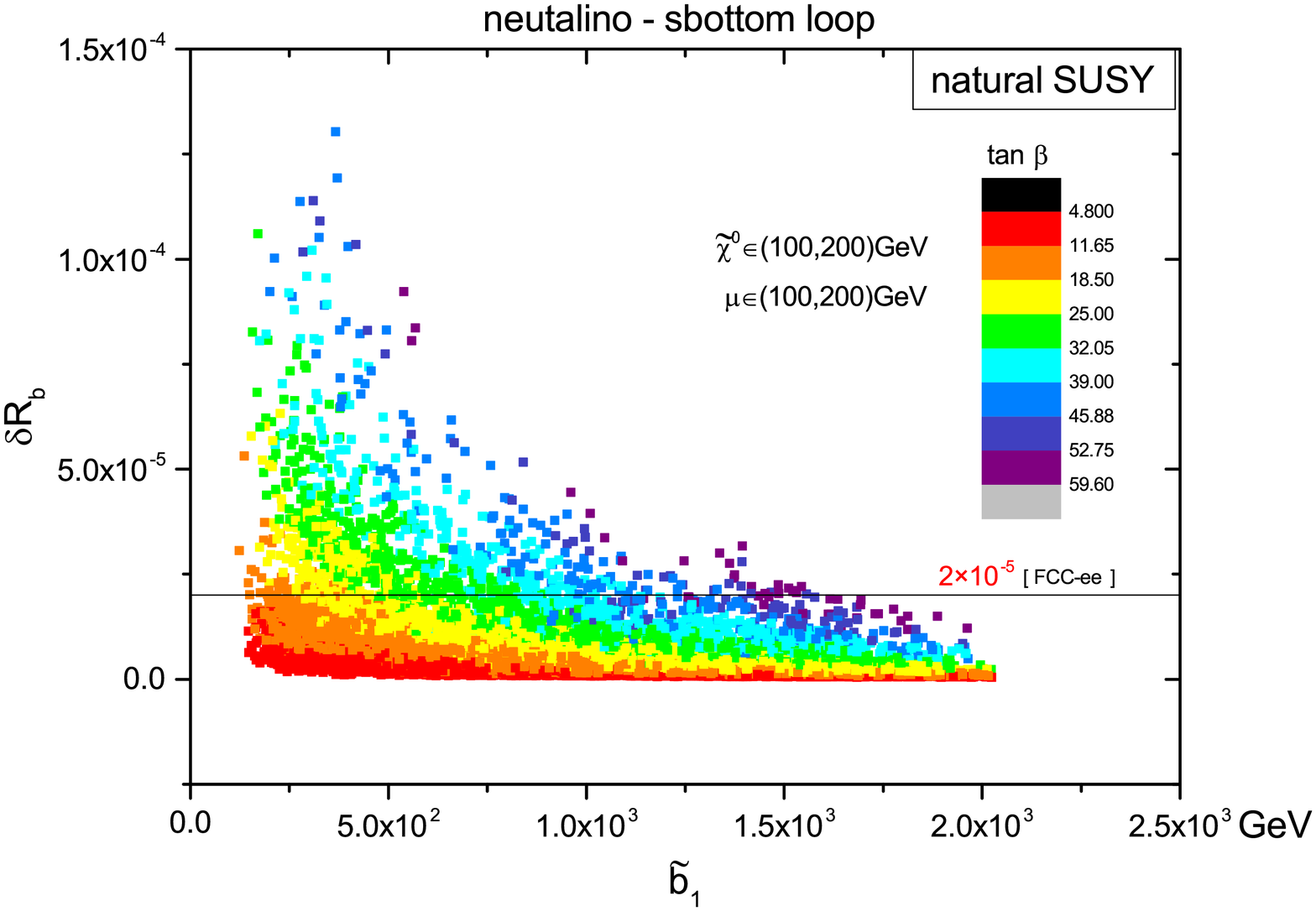}
   \vspace{-1.0cm}
    \caption{Same as Fig.\ref{dia:chargino}, but showing the \protect\\ neutralino loop effects.}
    \label{dia:neturalino}
  \end{minipage}%
  \begin{minipage}[t]{0.5\linewidth}
    \centering
    \includegraphics[width=1.0\textwidth,clip]{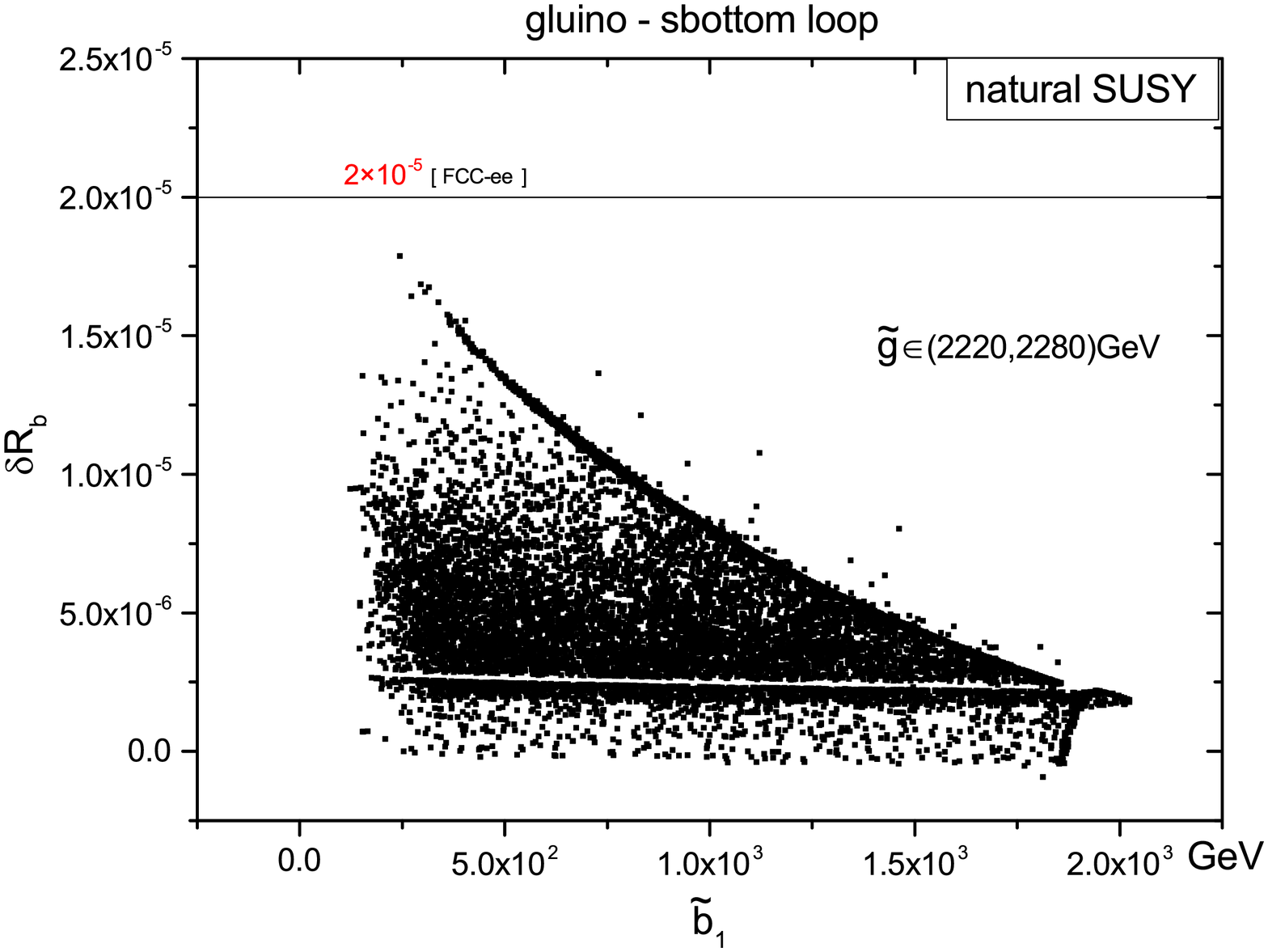}
   \vspace{-1.0cm}
    \caption{Same as Fig.\ref{dia:chargino}, but showing the \protect\\ gluino loop effects.}
    \label{dia:gluino}
  \end{minipage}
\end{figure}

\begin{figure}
  \begin{minipage}[t]{0.5\linewidth}
    \centering
    \includegraphics[width=1.0\textwidth,clip]{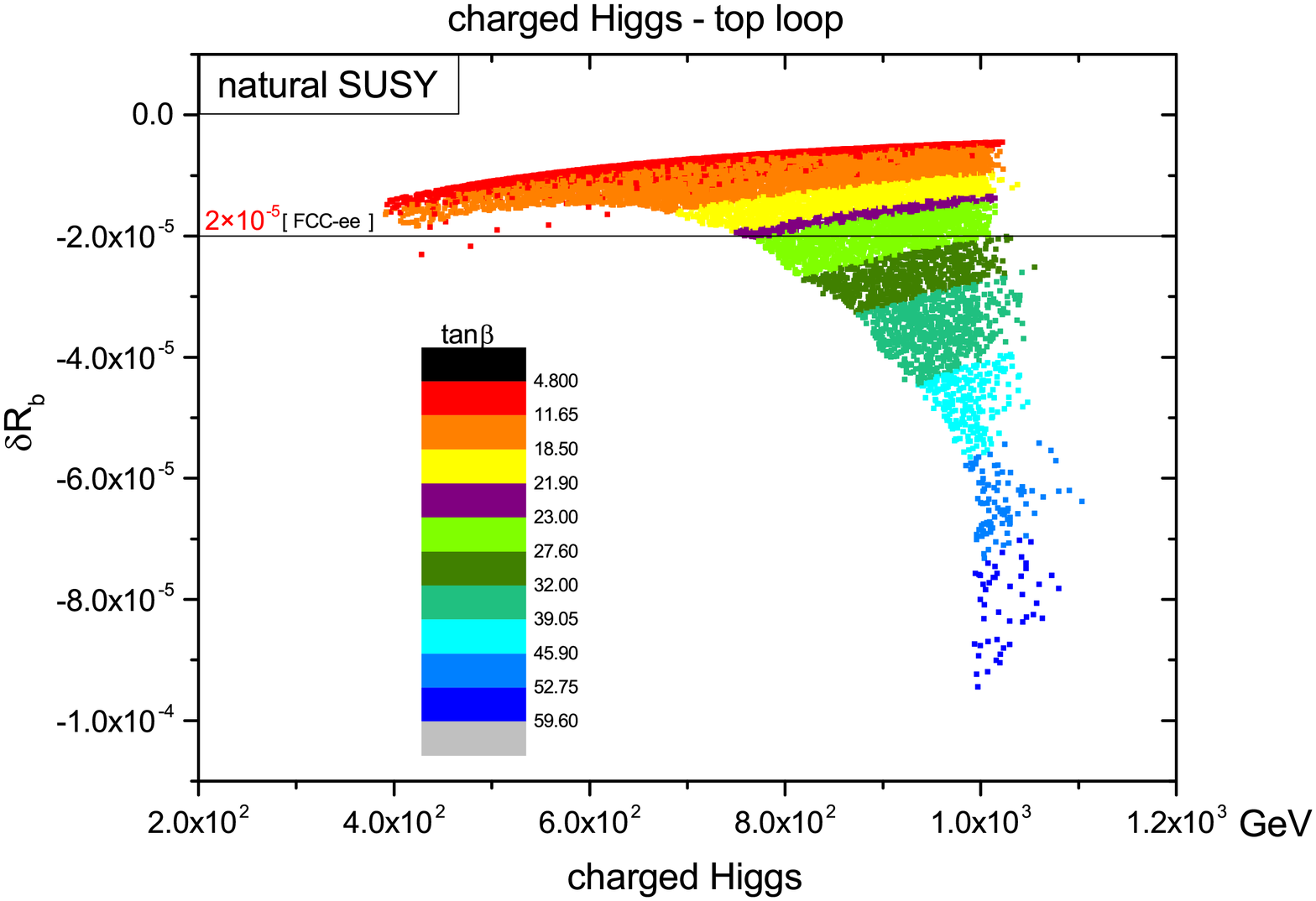}
   \vspace{-1.0cm}
    \caption{Same as Fig.\ref{dia:chargino}, but showing the \protect\\ charged Higgs loop effects.}
    \label{dia:chargedhiggs}
  \end{minipage}%
  \begin{minipage}[t]{0.5\linewidth}
    \centering
    \includegraphics[width=1.0\textwidth,clip]{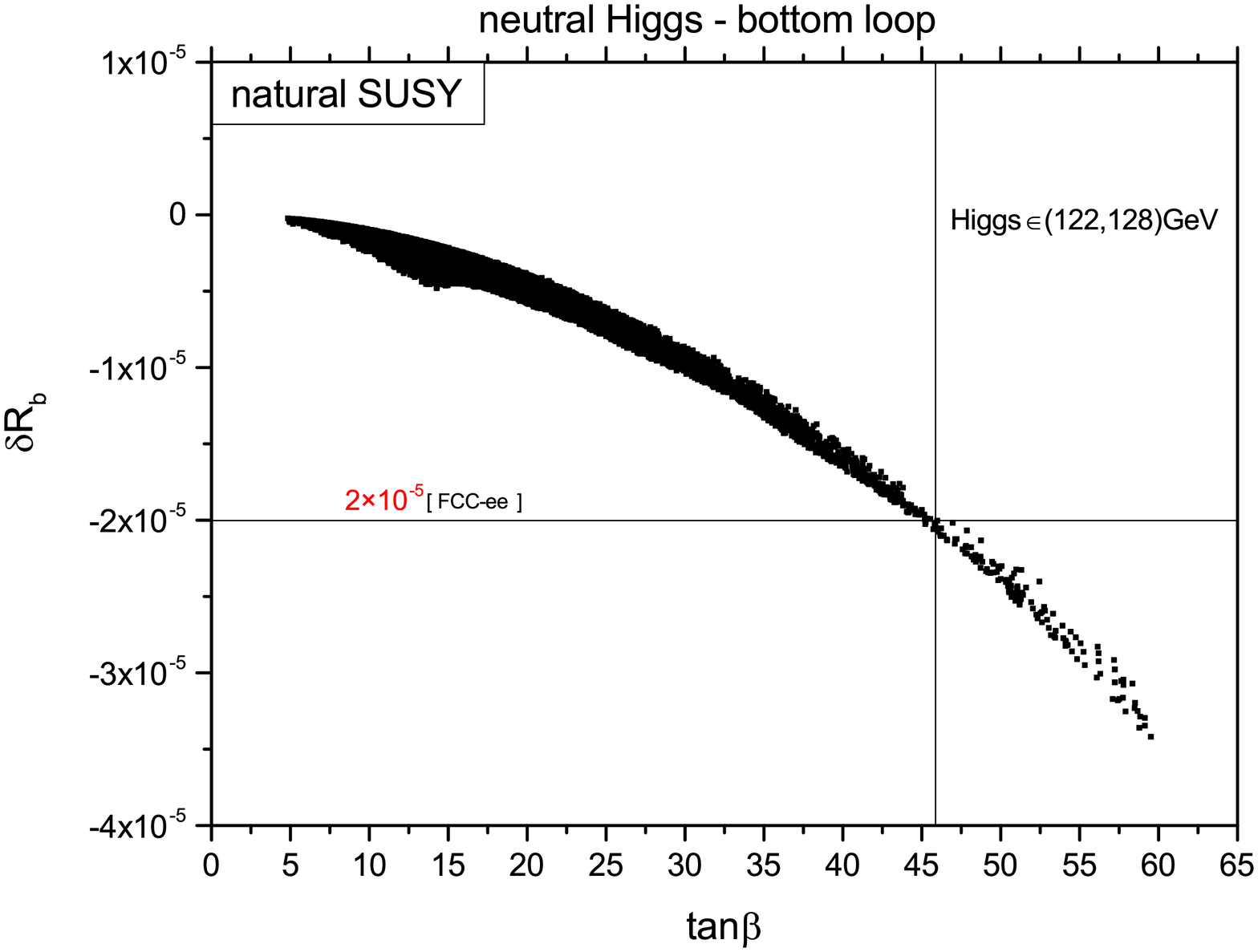}
   \vspace{-1.0cm}
    \caption{Same as Fig.\ref{dia:chargino}, but showing the \protect\\ neutral Higgs loop effects.}
    \label{dia:neutralhiggs}
  \end{minipage}
\end{figure}

\begin{figure}
  \begin{minipage}[t]{0.5\linewidth}
    \centering
    \includegraphics[width=1.0\textwidth,clip]{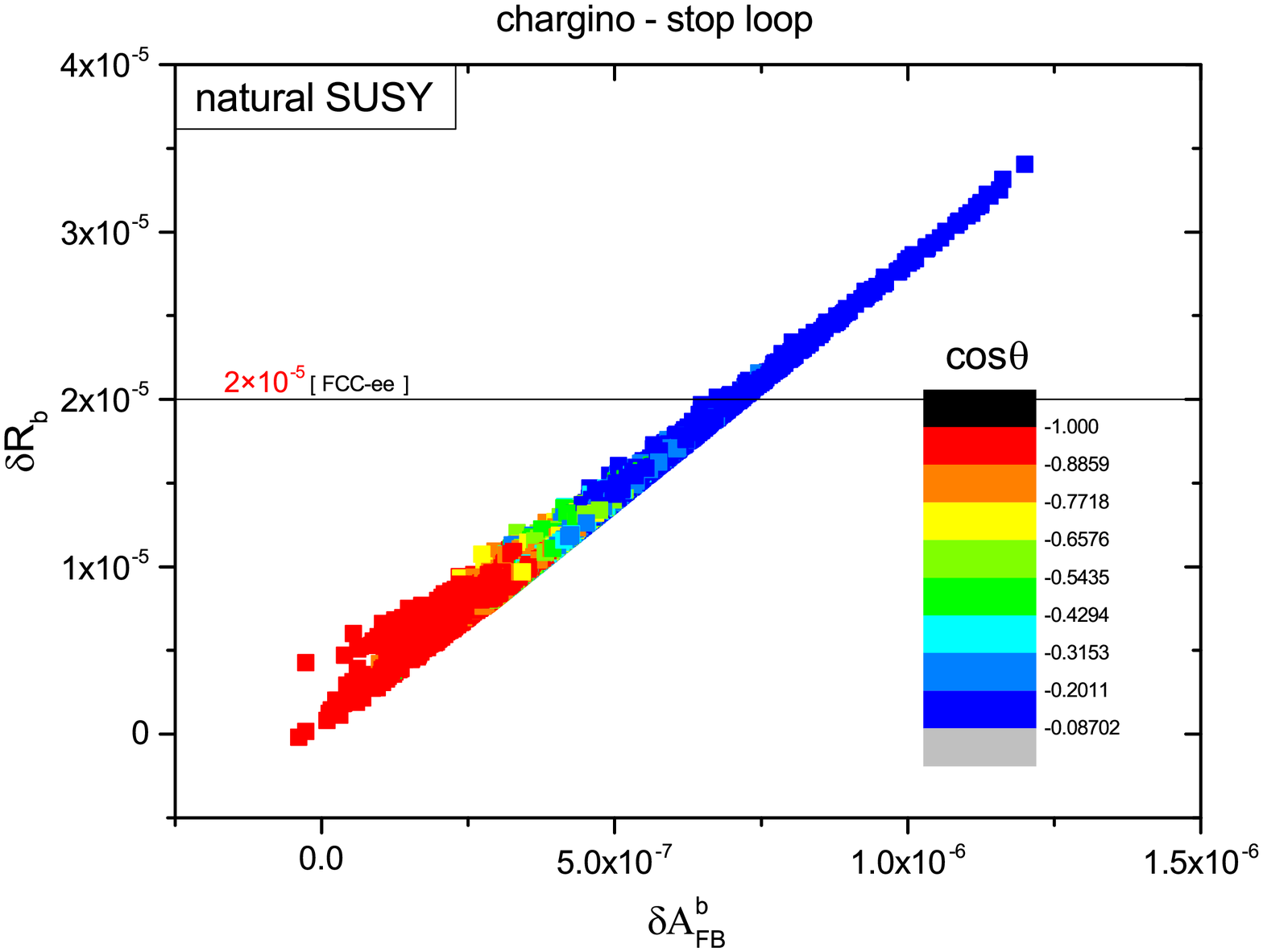}
   \vspace{-1.0cm}
    \caption{Same as Fig.\ref{dia:chargino}, but showing $\delta R_b$ \protect\\ versus $\delta A^{b}_{FB}$.}
    \label{dia:chargino-stop-dabfb}
  \end{minipage}%
  \begin{minipage}[t]{0.5\linewidth}
    \centering
    \includegraphics[width=1.0\textwidth,clip]{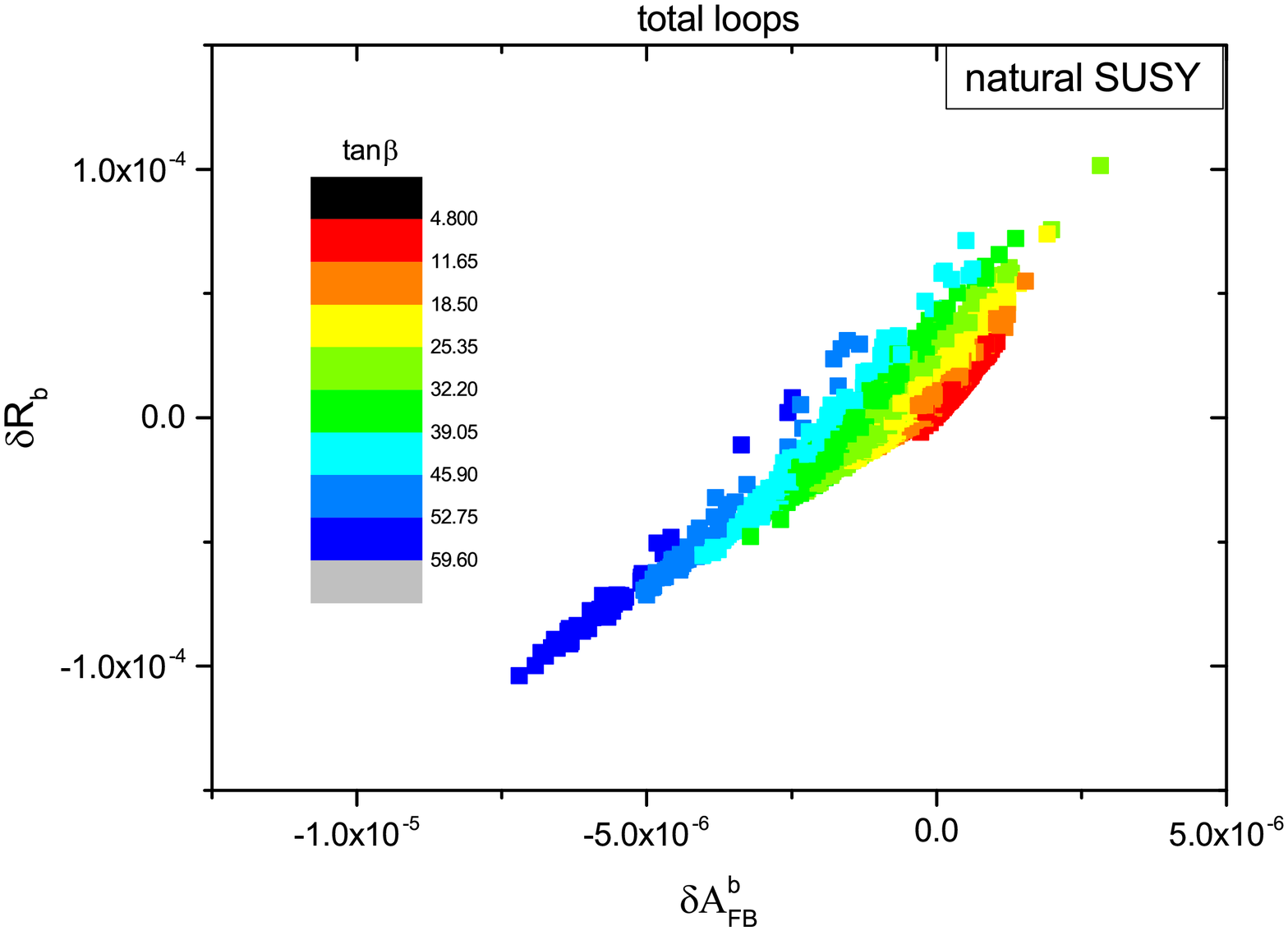}
   \vspace{-1.0cm}
    \caption{Same as Fig.\ref{dia:chargino-stop-dabfb}, but for the combined \protect\\ loop effects.}
    \label{dia:total-tan-dabfb}
  \end{minipage}
\end{figure}

\begin{figure}
  \centering
  \includegraphics[width=0.5\textwidth]{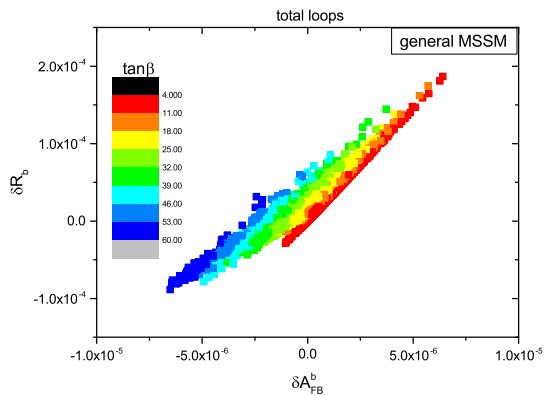}
  \vspace{-0.5cm}
  \caption{Same as Fig.\ref{dia:total-tan-dabfb}, but for the general MSSM.}
  \label{dia:total-tan-dabfb-general}
\end{figure}

\begin{figure}
  \centering
  \includegraphics[width=0.45\textwidth]{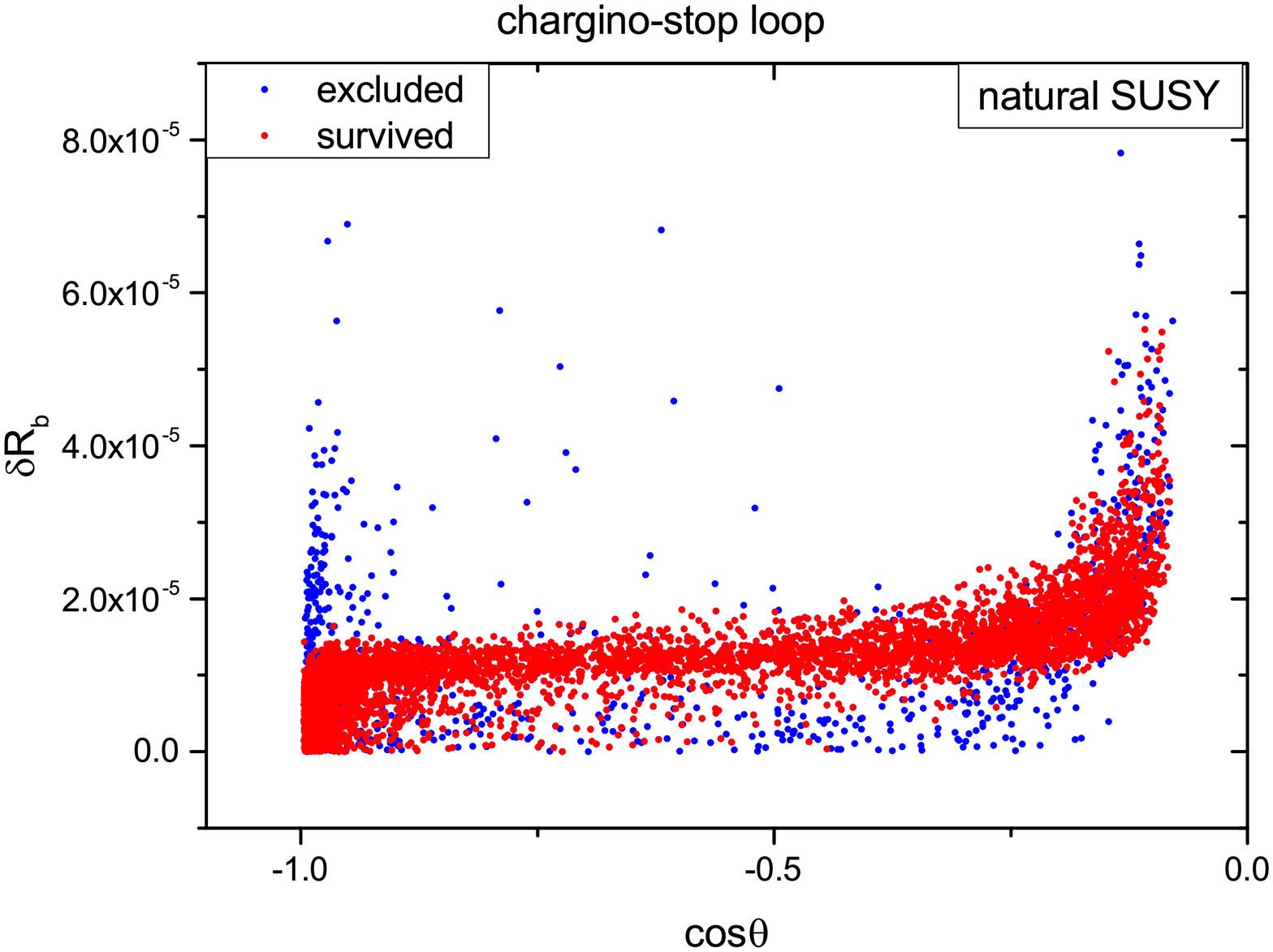}
  \includegraphics[width=0.45\textwidth]{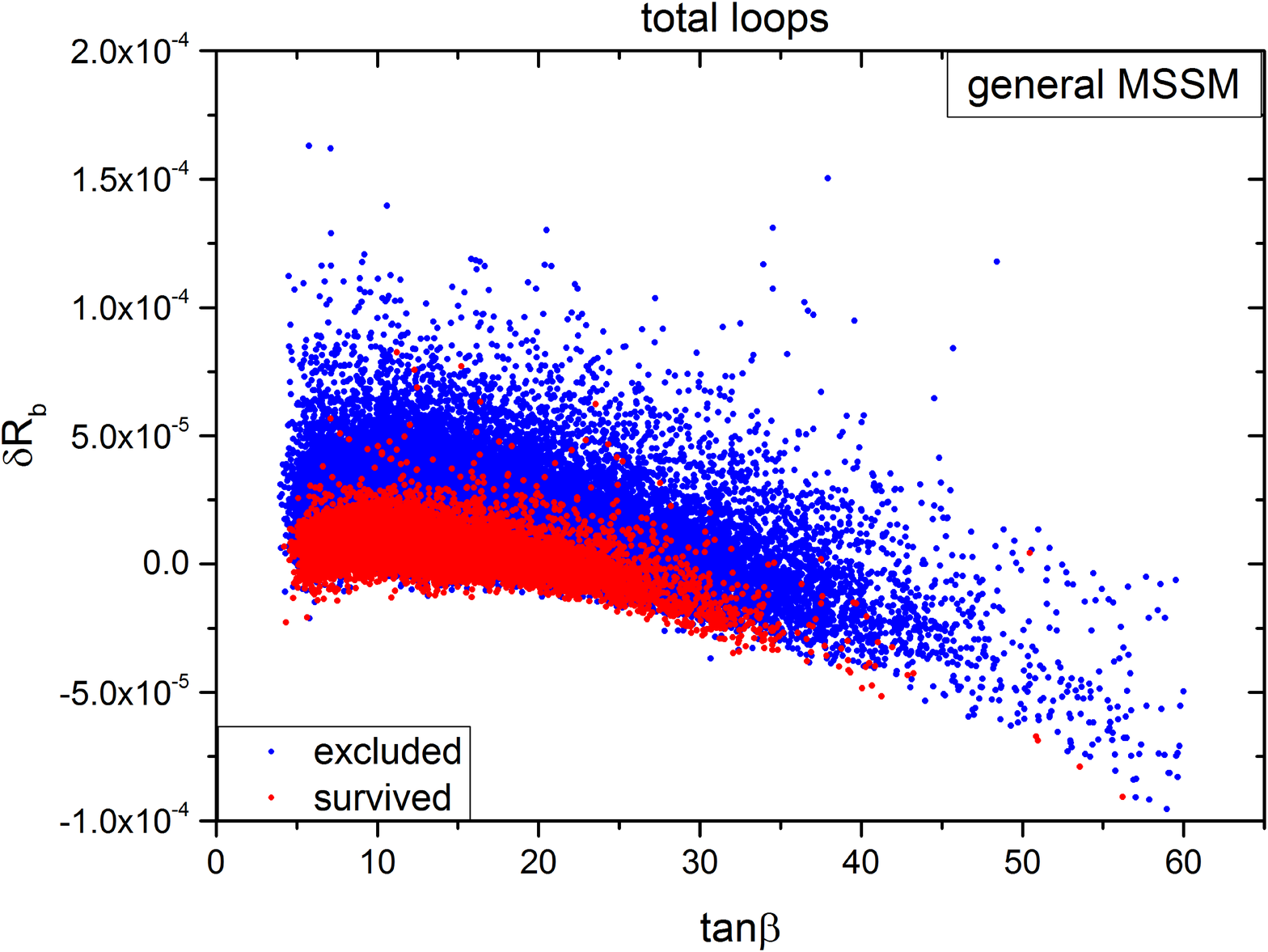}
   \vspace{-0.5cm}
  \caption{The plots of survived samples, showing
the most sensitive restrictions in two scenarios.
The left panel is for natural SUSY, where the samples
with and without B-physics constraints are displayed
(other constraints are satisfied).
The right panel is for the general MSSM, where the samples
with and without the dark matter-nucleon spin-independent scattering
limits are displayed (other constraints are satisfied).}
  \label{dia:restriction}
\end{figure}

The results for natural-SUSY and the general MSSM
are displayed in Figs.\ref{dia:chargino}-\ref{dia:total-tan-dabfb}
and Figs.\ref{dia:total-tan-dabfb-general}-\ref{dia:restriction}, respectively.
We first show the results of different loops and then show the combined results.
Finally we compare the natural-SUSY results with the general MSSM results.

About the future precision of $R_b$ measurement,
the CEPC would produce $10^{10}$ $Z$-bosons and probably measure $R_b$ with an uncertainty
of $1.7\times10^{-4}$ \cite{Fan:2014axa,CEPC}, while the FCC-ee could produce $10^{12}$ $Z$-bosons
and give a much better $R_b$ measurement at $10^{-5}$ level \cite{FCC-ee}.
In our figures, for illustration, we mark an uncertainty of $2\times10^{-5}$~\cite{FCC-ee,Fan:2014axa}. The SUSY parameter space
giving $\delta R_b^{\rm SUSY} > 2\times10^{-5}$ corresponds to the observable region.


Some discussions about the results are in order:
\begin{itemize}
\item[(a)]
From Fig.\ref{dia:chargino} we see that the chargino-stop loop effects are sizable
only if $\tilde t_1$ is dominated by a right-handed stop.
For a left-handed stop, its coupling with higgsino and bottom $Y_b = g m_b/(\sqrt{2} m_W \cos\beta)$
is suppressed (the lightest chargino $\tilde\chi^\pm_1$ is dominated by higgsino component since
the higgsino mass $\mu$ is much smaller than the gaugino masses $M_1$ and $M_2$ in natural SUSY).
Only for a very large $\tan\beta$ can the coupling $Y_b$ be comparable to the corresponding
right-handed stop coupling $Y_t=g m_t/(\sqrt{2} m_W \sin\beta)$.  Our numerical results shows
that $\tan\beta$ is smaller than 35 (so that $Y_b/Y_t<1$) for $\tilde t_1$ below 530 GeV
(when $\tan\beta$ is larger, $\tilde t_1$ must be heavier to satisfy the experimental constraints).
Note that, as commented in the preceding section,  so far the right-handed stop mass
in natural SUSY is weakly bounded by LHC experiments (its mass can be as light as 210 GeV
for higgsinos heavier than 190 GeV) \cite{stop-bound}.

\item[(b)] As shown in Fig.\ref{dia:gluino}, the gluino-sbottom loop effects are
very small due to the heaviness of gluino.
The loop effects of the neutralinos, charged and neutral Higgs bosons, as shown in
Figs.\ref{dia:neturalino}, \ref{dia:chargedhiggs} and \ref{dia:neutralhiggs},
are sensitive to $tan\beta$ and can be sizable for a large value of $\text{tan}\beta$. Our numerical results show that the neutralino loop can push the $\tilde b_1$ mass to 850 GeV when $\text{tan}\beta$ is around 32. If the $\text{tan}\beta$ is about 23, through the charged Higgs loop, $\tilde H^+$ mass less than 770 GeV is excluded. The neutral Higgs loops impose an upper bound of 46 on
the value of $\text{tan}\beta$.

\item[(c)] From Figs.\ref{dia:chargino-stop-dabfb}, \ref{dia:total-tan-dabfb}
and~\ref{dia:total-tan-dabfb-general} we see that
the SUSY effects in  $R_b$ and  $A^{b}_{FB}$ are correlated, as expected.
Both observables can jointly probe the SUSY effects. While the chargino loop effects
always enhance both quantities, the combined total effects of all loops can either enhance
or reduce them. We also find that in the general MSSM without special naturalness requirement,
both $R_b$ and  $A^{b}_{FB}$ are allowed to vary in a larger region than in natural SUSY,
especially when $\tan \beta$ is small.

\item[(d)] From Figs.\ref{dia:chargino}-\ref{dia:total-tan-dabfb-general} we see that
in some currently allowed parameter space, the effects of natural SUSY may be accessible
in the future $R_b$ measurement. If it can be measured with an uncertainty of $2\times10^{-5}$,
a large part of SUSY parameter space can be covered.

\item[(e)]
We found that for natural SUSY the most stringent limits are from B-physics,
while for the general MSSM the most stringent limits are
from the dark matter-nucleon spin-independent scattering limits.
The results are shown in Fig.\ref{dia:restriction}.
Other constraints, such as the dark matter-nucleon spin-dependent scattering
cross section, are also making impacts but not as stringent as these two.

\end{itemize}

\section{Conclusion}
\label{sec:conclusion}
We revisited the SUSY effects in $R_b$ under
current experimental constraints including the LHC Higgs data, the $B$-physics
measurements, the dark matter relic density and direct detection limits, as well as
the precision electroweak data. We scanned over the SUSY parameter space and
in the allowed  parameter space we displayed the SUSY effects in $R_b$.
We found that although the SUSY parameter space has been severely
restrained by current experimental data, SUSY can still alter $R_b$ with a
magnitude sizable enough to be observed at future $Z$-factories (ILC, CEPC, FCC-ee).
Assuming a precise measurement $\delta R_b = 2.0 \times 10^{-5}$ at FCC-ee,
we can probe the right-handed stop to 530 GeV through the chargino-stop loops,
probe the sbottom to 850 GeV through the neutralino-sbottom loops and
the charged Higgs to 770 GeV through the Higgs-top quark loops for a large
$\text{tan}\beta$.
The full one-loop SUSY correction to $R_b$ can reach $1 \times 10^{-4}$
in natural SUSY and $2 \times 10^{-4}$ in the general MSSM.

\acknowledgments
We would like to thank Junjie Cao, Lei Wu,
Yang Zhang, Mengchao Zhang for useful discussions.
This work has been supported in part by the National
Natural Science Foundation of China under grant Nos. 11275245 and 11135003,
and by the CAS Center for Excellence in Particle Physics (CCEPP).

\end{spacing}
\end{document}